# Changes in fractal properties of geomagnetic indexes as possible magnetic storms precursors

Vitor H. A. Dias[1,2], Jorge O. O. Franco[1,2] and Andrés R. R. Papa[1,2,*]

[1]Observatório Nacional, Rua General José Cristino 77, São Cristóvão,
Rio de Janeiro, 20921-400 RJ, BRASIL
[2]Instituto de Física, Universidade do Estado do Rio de Janeiro, Rua São Francisco Xavier 524, Maracanã,
Rio de Janeiro, 20550-900 RJ, BRASIL

**Abstract**

Records of Dst, Kp and Sym–H indexes are analyzed looking for evidences of possible precursors for magnetic storms. For this the main magnetic storms were located and some periods immediately before storms and some periods well before them studied. Statistical properties of both types of periods were then compared. In particular, were compared the slopes of the power laws that have been found for the distributions of index values. A systematic deviation was found between both distributions for the Kp index. It was not so for Dst and Sym-H indexes. For the three indexes it was found a correlation between slopes and the corresponding storm intensity, which could serve as a probabilistic approach to magnetic storms forecasting. The data for the analysis was obtained at the World Data Centre for Geomagnetism at Kyoto.

___________________________

* E-mail address: papa@on.br

**Introduction**

Geomagnetic storms are (together with geomagnetic reversals [*Ponte-Neto and Papa*, 2006]) the most striking phenomena in the magnetic field that we can measure at the Earth's surface. They are periods from one to three days during which the magnetic field born at the Sun surface and atmosphere of the Earth presents strong variations. Those variations seriously affect telecommunication transmissions. There are some evidences that during magnetic storms the number of heart attacks increases in relation to calm periods [*Halberg et al.*, 2001]. During the year of 1989 the region of Quebec, in Canada, suffered a total energy blackout for more than nine hours caused by a strong magnetic storm. Geomagnetic storms may also affect the development of other technical and commercial activities as, for example, measurements of crustal magnetic fields for prospecting and directional drilling [*Gleisner et al.*, 2005]. All these reasons justify the search for methods that allow predict magnetic storms with enough antecedence. But as in the case of earthquakes and other catastrophic events that affect human activities (or even, the human life) there do not exist yet methods to accomplish this task in a reliable manner.

During the last five decades many efforts have been devoted to study statistical properties of magnetic measurements at the Earth's surface and satellites. We can mention, among many others, a power spectrum for the magnetic field around the world [*Alldredge et al.*, 1963], and more recently, a detrended fluctuation analysis for the Sym-H index [*Wanliss*, 2005], an intermittency analysis of geomagnetic storm series [*Bolzan et al.*, 2005] and a

Fourier filtering procedure to study statistical signatures of direct magnetic field measurements [*Papa et al.*, 2006].

The search for geomagnetic storms precursors range from some Earth grounded systems as, for example, muon detectors [*Jansen et al.*, 2001], to some Earth and satellites studies on visible light and X-ray burst during solar flares [*Yermolaev and Yermolaev*, 2003]. However, there are some discrepancies on the effective relation between solar flares and geomagnetic storms [*Yermolaev and Yermolaev*, 2006] and this is a point of current discussion. There is a comprehensible review [*Lathuillère et al.*, 2002] on scientific models for space weather developments. We have not found works searching for precursor using the fractal properties of actual geomagnetic measurements or their related indexes.

On the other hand, magnetic storms are themselves possible candidates to precursors of other geophysical phenomena like earthquakes [*Enescu*, 2005; *Kushwah*, 2004] but as well as in the case of flare-storms relations this possibility requires further scrutiny.

When preparing this work we were mostly inspired on the works by *Wanliss* [2005] and by *Papa et al.* [2006]. However our work follows well defined different lines. *Wanliss* [2005] concentrated on an extensive study (from 1981 to 2000) of fractal properties of the Sym-H index dividing the record in quiet (periods of at least 1000 minutes of Kp less or equal to 1) and active intervals (periods of at least 10000 minutes of Kp greater or equal to 4). On the other hand, *Papa et al.* [2006] concentrated their efforts in studying a short period of time (October 2000) through a Fourier analysis of magnetic measurements that overcome some of the problems that emerge because the non-stationary character of those series. None of them focused its attention in the search for magnetic storms precursors. Here we are interested in compare fractal properties but only of relatively calm periods, some of them near magnetic storms' beginnings and some others far (see below) from magnetic storms.

**Data analysis and Preparation**

The 1-hour-resolution Dst index, which is essentially the same as the minute Sym-H index except for the time resolution, is usually taken as the magnetic storm revealer. It is often taken the value –50nT of that index as the threshold for considering that a magnetic storm is taken place [*Maltsev*, 2003]. In Figure 1 we present the variation of the Dst index for the month of October 2000. The data was obtained from the World Data Centre for Geomagnetism at Kyoto (WDC). Analyzing Figure 1 and other one-month periods (not shown) it is apparent that the unique general feature for the series is that does not exist any general feature for it. However, there are two characteristics that are present in a considerable number of cases: a sudden decrease in Dst values (during a few hours) and a slow recover to values near zero during a period of one to three days. On those features was based our data preparation.

It is well known that the Dst series present a highly non-stationary character. To avoid problems resulting from this fact there are two customary policies that are implemented. The first studies long periods of time searching for robust averages [*Wanliss*, 2005]. The second studies short periods of time, where a quasi-stationary character can be assumed. It was our choice the last one.

The preparation of data consists in a set of relatively simple steps that we describe now. We have used as threshold criteria for the beginning of a magnetic storm a value of –50nT for the Dst index. With this, we have located the date and hour of the beginning of all the magnetic storms in the period 1998–2002. The beginning of the transient before storms was estimated to be the last value greater or equal to –10nT before the value less or equal to –50nT that marks the storm beginning. We have not considered transient periods greater than 10 hours and the corresponding "storm" does not enter our record. They were considered walking of Dst in far from zero values or rests of previous storms. The mean value for those periods corresponds to around 6 hours, and this value was taken as unique for all the storms. It is clear that not all the periods with values greater or equal to –50nT classify as storm with those criteria. We have also eliminated from our record the periods of Dst less than –50nT too close to previous storms (because we can not have a "clean" six days plus six hours period for our calculations, see below). The final result was that we eliminated around 50% of the periods with some point at Dst values lower than –50nT, giving 60 magnetic storms. Table 1 presents the set of storms that we have considered. We have investigated then two time windows of three days: the three consecutive days before the transient beginning and the three consecutive days before them. On both types of periods we have implemented a fluctuation analysis. Three days seems to be a reasonable period for storms forecasting [*Gleisner et al.*, 2005].

Figure 2 shows the distributions of Dst, Kp and Sym-H values for a month period (October 2000). All of them follow power laws with a given slope. The change of slope from one type of period to the other will be the focus of our attention. We have calculated the distribution function for Dst, Kp and Sym-H values for each of those three days periods. Note however that this is a quite dangerous strategy because we have small samples and, consequently, low quality power laws. The analysis was done in periods of three days while in Figure 2 the plots are for one month. These means that, instead of 240 data points for Kp, for example, there are only 24 points, and similarly for the other two indexes. Among all of them Kp is the more critical one (because is the temporarily less dense index). This was an extra motive to a further reduction in the set of magnetic storms to be considered for each of the indexes. Once calculated the slope and its error for both periods corresponding to a single storm it was established a threshold for the error. When the error was above this threshold the corresponding storm was dropped out of the initial set of 60 in Table 1, for that index. The thresholds for Dst, Kp and Sym-H indexes were 30%, 50% and 30%, respectively. With this the number of effectively used storms was reduced to 39 for Dst, 26 for Kp and 43 for Sym-H.

**Results and Discussion**

It is well known that the magnetic field measurement distributions [*Papa et al.*, 2006] as well as some index distributions [*Wanliss*, 2005] present power laws or, in other words, are compatible with some fractal system. Power laws have the form:

$$f(x) = k \cdot x^d \tag{1}$$

where f(x) is the frequency distribution on the variable x, k is a proportionality constant and d is the exponent of the power law.

Figure 3 shows the slope *d* distribution for Dst, Kp and Sym-H indexes. For Dst, as well as for Sym-H, there are no statistically relevant differences between the distributions for periods near and far from storms. For Kp it is apparent a trend to lower values of the power law exponent distribution when we consider periods closer to storms. We expected the results for Dst and Sym-H be similar because they are essentially the same except for the time resolution.

That there are not relevant differences (the difference in Kp must be caused by the low statistics) between both types of periods could be expected from previous results [*Wanliss*, 2005; *Papa et al.*, 2006] on the fractal character of those distributions: they are, in principle, indistinguishable.

The dependence on time of the slope for the whole period 1998-2002 is shown in Figure 4 for the three indexes. While there is not a one-to-one correspondence between the slopes of periods corresponding to a single storm, it can be noted that there is a global trend in Dst (increasing with time) and Sym-H (decreasing as time goes by) indexes. For Kp it was not possible to establish a unique trend for both dependences.

In Figure 5 we present the correlation between the slopes for both periods (near and far). There is not an apparent correlation for none of the indexes. However, there seems to be "forbidden areas" in all of them. For example, in Figure 5b, for Kp, a d (far from storm) value lower than –0.4 excludes the occurrence of values lower than –0.5 for d (near storm).

The analysis presented so far does not involve the possibility of storms prediction. It was more oriented to characterize changes in the statistical properties of indexes from one type of period to the other. In Figure 5 it is shown the dependence of d for both periods on the corresponding storm intensity. There are forbidden regions in all the cases. When we try to predict some phenomenon forbidden regions are important. Although they do not say what is going to happen, they tell us what is not going to occur. They can also provide us with the probability that a given event happen. Let us briefly exemplify both affirmations. From Figure 6a for index Dst we could say that if the exponent of the power law (for any of the periods) is lower than –3.5 then the next storm will be no more intense than –125nT. From Figure 6a we can also extract that if the exponent (for a near period) is between –2.5 and 0 there is a probability of $6/21 \cong 30\%$ for the next storm of being more intense than –125nT. The quantity 6/21 was obtained by noting that there are 6 points in the rectangle formed by the values d = –2.5 and 0 and storm intensity = –325nT and –125nT, and 21 points in the rectangle formed by the same values for d and storm intensity = –125nT and –25nT.

It should be noted in Figure 6 that forbidden regions delimited by points corresponding to periods near storms (circles) are larger than those delimited by points corresponding to periods far from storms. They impose more restrictive conditions. This is intuitively clear, if we accept, as it seems to be, that there is some relevant information on the storms in periods before them. The closer the analyzed period to the storm the more reliable information we have on it. The existence of correlations between slopes and storm intensities means that, although the system has a partial fractal character, there are components of other types of phenomena that allow this correlation.

We consider the forbidden regions in *d* versus storm-intensity plots our main result.

**Conclusions**

We have studied the change in fractal properties of magnetic indexes of wide use. There are several things to be learnt from our study. Our mean time for transients before storms is a first order approximation (note that the variation ranges from 3 to 10 hours). Probably a case-by-case study will give more reliable results. Our studies do not included some storms close to previous ones. A possible way to include those cases would be to implement a similar procedure in the long tail part of the storms. Multi-peak storm stayed also out of our scope and we have no idea on how to extend our study to include them. They seem to be stresses not completely relaxed during the main phase (first peak) of the storm. In any case, further and more detailed studies will be necessary. A better statistics (that in our case means longer periods of time) would be necessary. The inverse way of validation is also mandatory. There are some works running in those directions and will be published elsewhere.

**Acknowledgements**

The authors sincerely acknowledge partial financial support from FAPERJ (Rio de Janeiro Founding Agency) and CNPq (Brazilian Founding Agency).

**Table Caption**

Table 1.- Storms following our classification. The column "Date" presents the dates when a value Dst ≤ –50nT was reached for the first time. The column "Depth" gives which this value was. The column "Transient" tell us the moment when the last value of Dst ≥ –10nT before the storm occurred. Finally, the column "Time" gives the time (UT) at which the storm started.

**Figure Captions**

Figure 1.- Dst dependence for a period of one month (October 2000). The data was obtained from the World Data Centre for Geomagnetism at Kyoto.

Figure 2.- Distribution of indexes values for the month October 2000. a) Dst, b) Kp and c) Sym-H. All the distributions follow reasonably well power laws in some interval. Note that for Dst and Sym-H indexes we have used a log-log plot while, given that Kp is itself a quasi-logarithmic index, for Kp a semi-log one. In all cases the bold line is a guide for the eye.

Figure 3.- Distributions of slopes (*d*) values for periods far from storms (squares) and near storms (circles). a) Dst, b) Kp and c) Sym-H. There seems to be a statistically relevant difference for the Kp index.

Figure 4.- Time (storm number) dependence of slope for both kinds of periods. a) Dst, b) Kp and c) Sym-H. There seems to be common long-range trends in Dst and Sym-H indexes.

Figure 5.- Slope-slope correlations for the three indexes. a) Dst, b) Kp and c) Sym-H. Some "forbidden" regions seem to be present.

Figure 6.- Slope – magnetic storm intensity dependence. a) Dst, b) Kp and c) Sym-H. Some "forbidden" regions seem to be present here also. In this case they can be tools for magnetic storm forecasting.

Figure 1

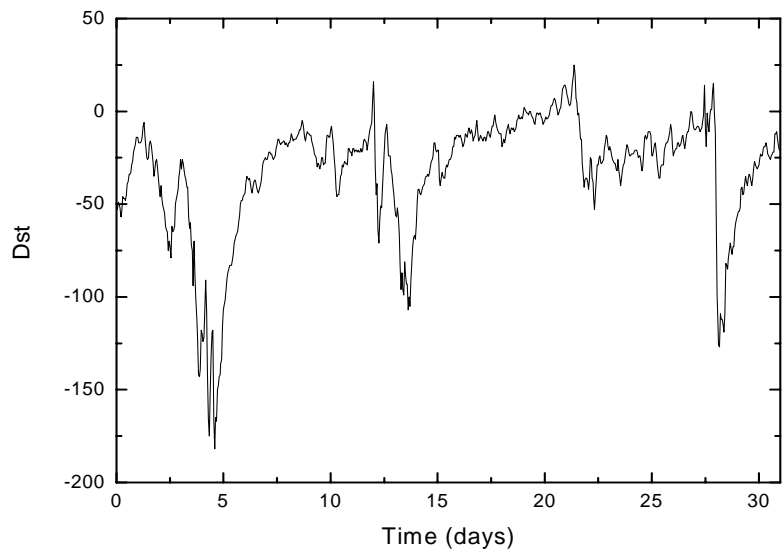

Figure 2

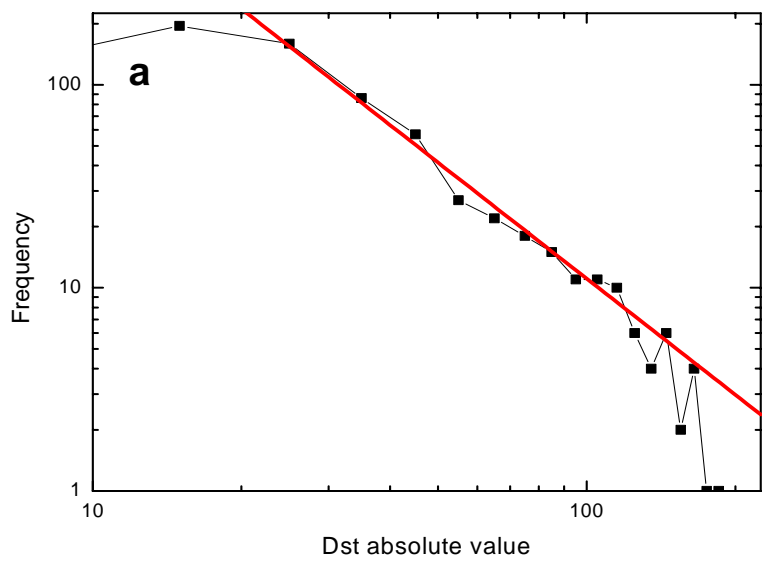

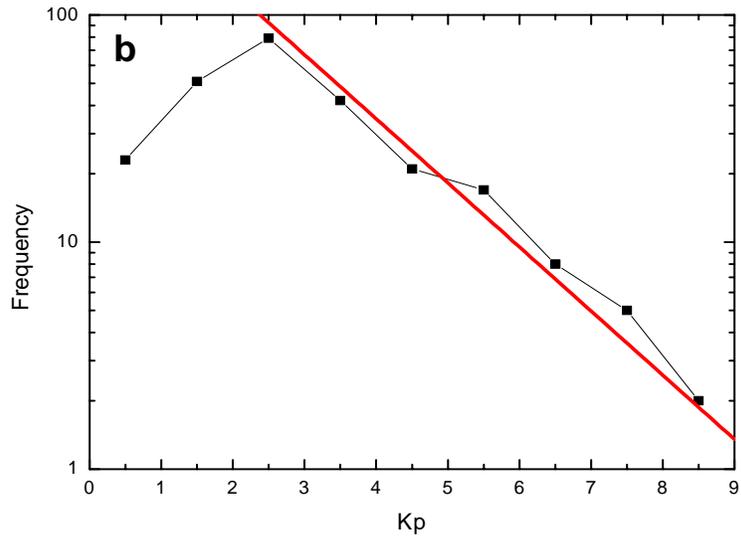

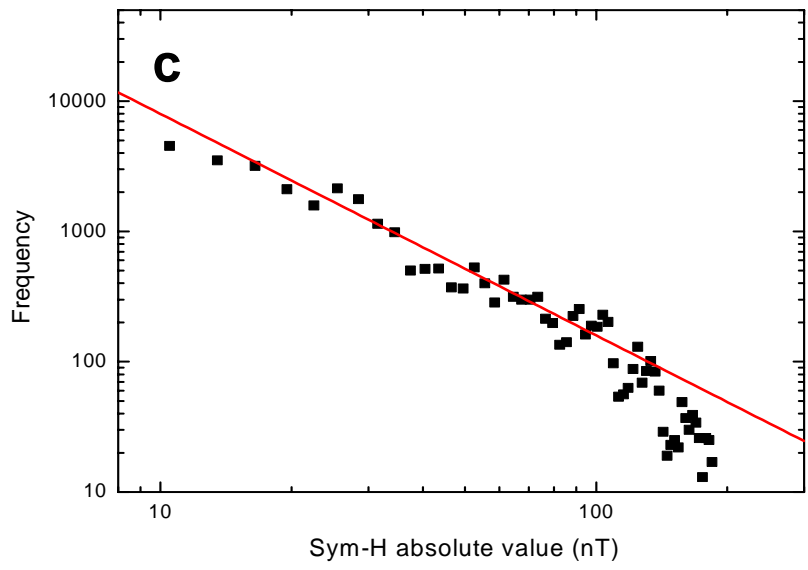

Figure 3

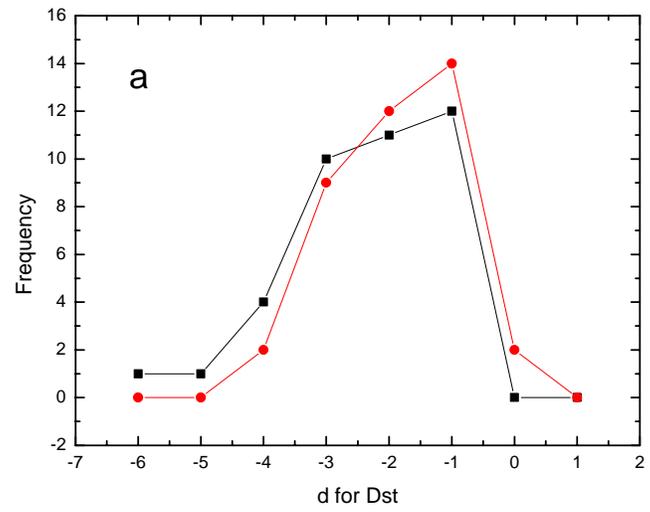

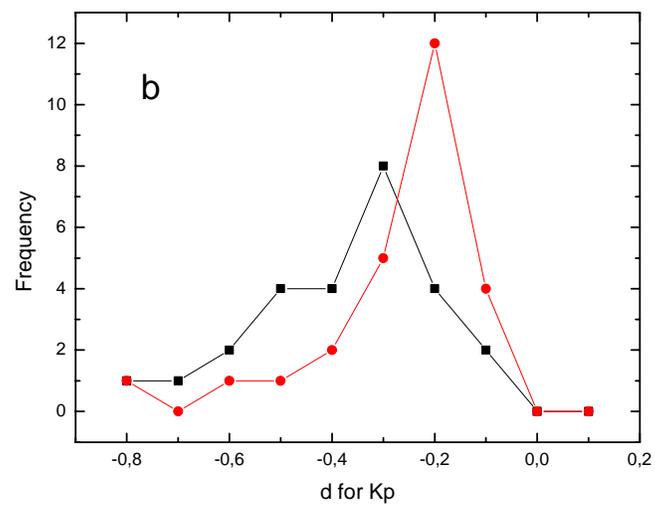

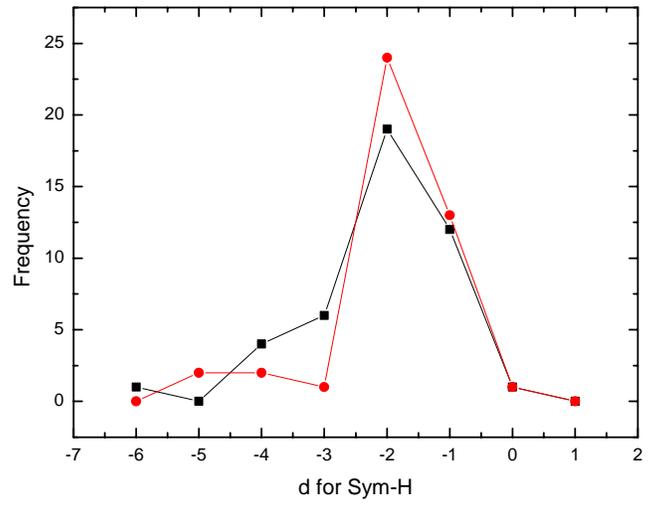

Figure 4

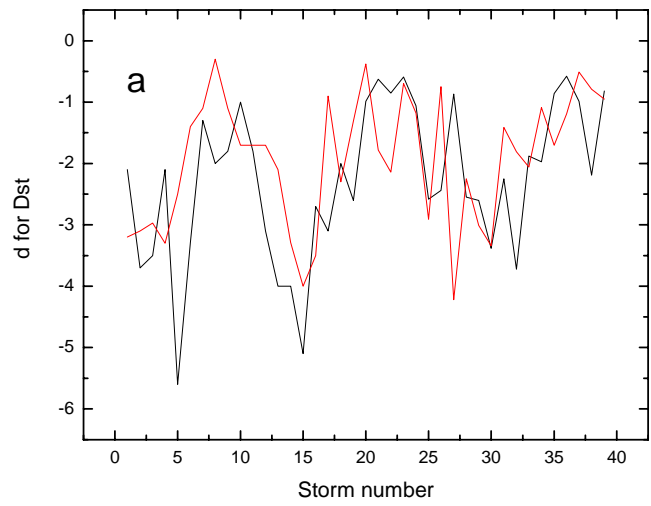

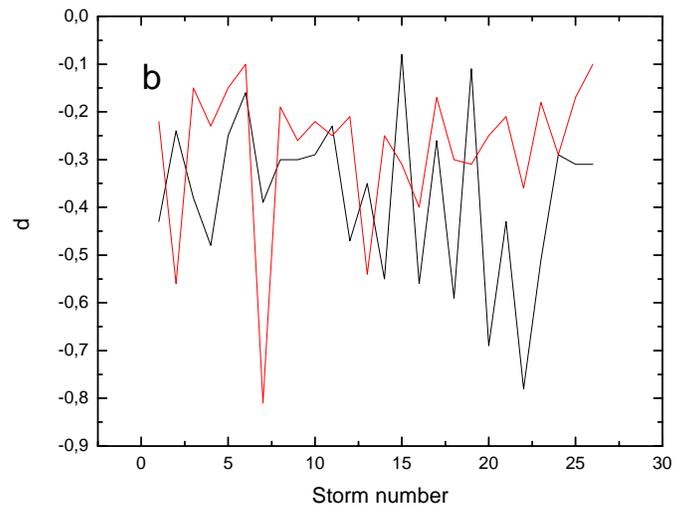

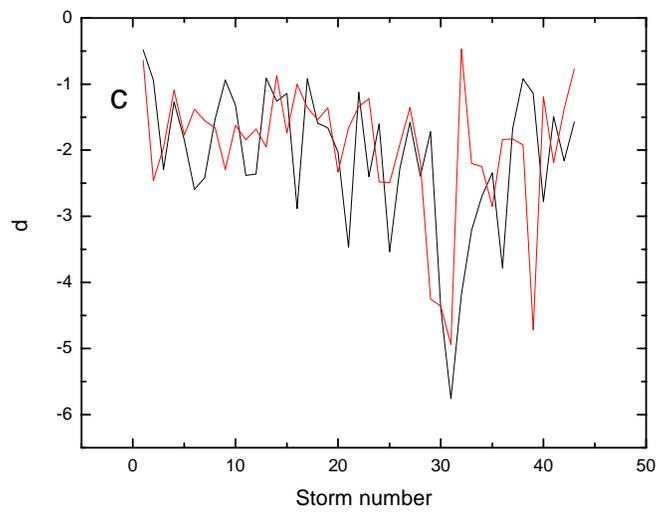

Figura 5

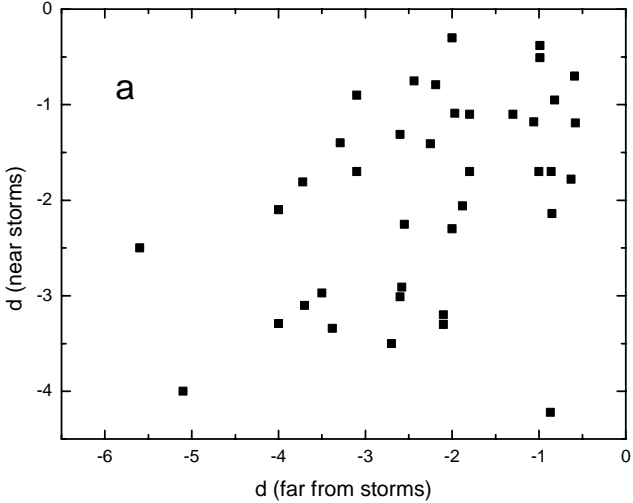

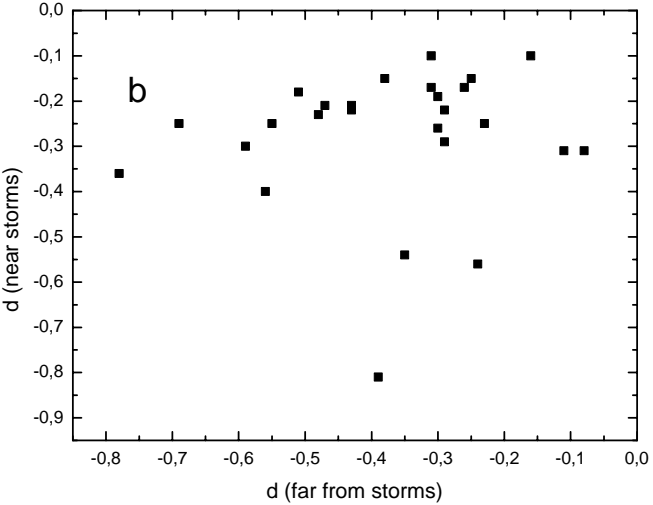

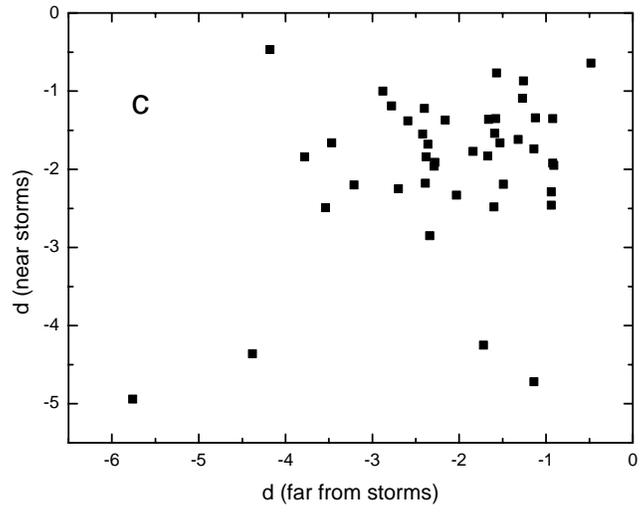

Figure 6

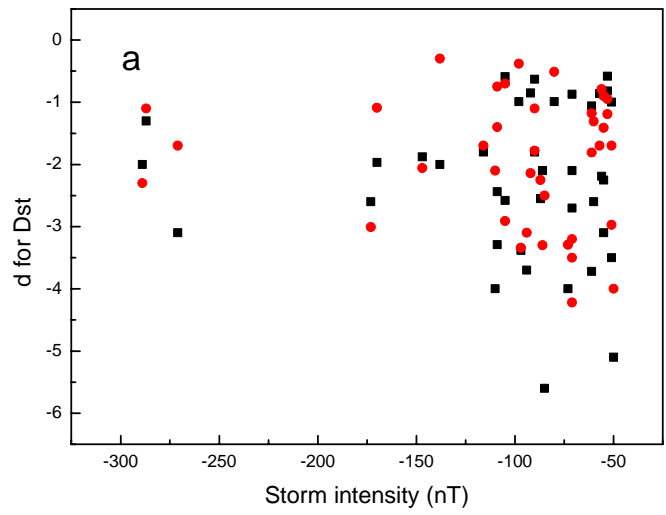

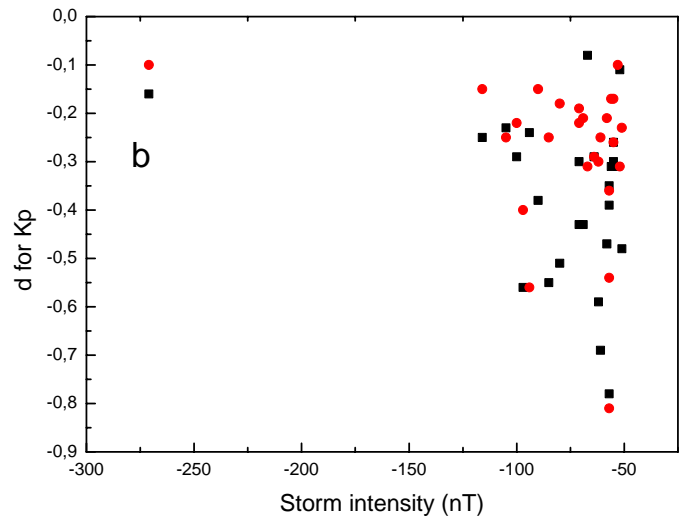

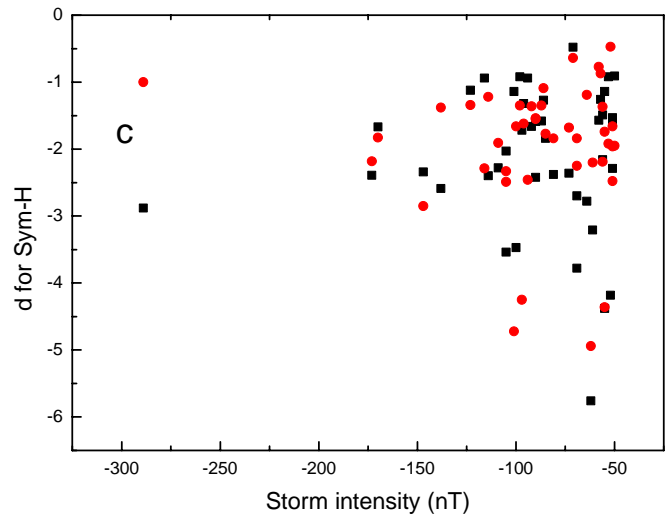

Table 1

| No. | Date (d/m/y) | Depth (NT) | Transient (hours) | Time (UT) | No. | Date (d/m/y) | Depth (nT) | Transient (hours) | Time (UT) |
|---|---|---|---|---|---|---|---|---|---|
| 1 | 18/02/98 | -100 | 10 | 01:00 | 32 | 13/10/00 | -71 | 5 | 06:00 |
| 2 | 10/03/98 | -116 | 7 | 21:00 | 33 | 10/11/00 | -96 | 5 | 13:00 |
| 3 | 21/03/98 | -85 | 6 | 16:00 | 34 | 27/11/00 | -80 | 3 | 02:00 |
| 4 | 24/04/98 | -69 | 7 | 08:00 | 35 | 23/12/00 | -62 | 6 | 05:00 |
| 5 | 02/05/98 | -85 | 8 | 18:00 | 36 | 13/02/01 | -50 | 4 | 22:00 |
| 6 | 14/06/98 | -55 | 6 | 11:00 | 37 | 19/03/01 | -105 | 9 | 22 |
| 7 | 26/06/98 | -101 | 4 | 05:00 | 38 | 28/03/01 | -56 | 4 | 02:00 |
| 8 | 06/08/98 | -138 | 8 | 12:00 | 39 | 11/04/01 | -271 | 8 | 24:00 |
| 9 | 18/09/98 | -51 | 4 | 14:00 | 40 | 18/04/01 | -114 | 5 | 07:00 |
| 10 | 25/09/98 | -170 | 4 | 05:00 | 41 | 18/06/01 | -61 | 4 | 09:00 |
| 11 | 19/10/98 | -109 | 10 | 13:00 | 42 | 17/08/01 | -105 | 4 | 22:00 |
| 12 | 25/12/98 | -57 | 4 | 12:00 | 43 | 13/09/01 | -57 | 5 | 08:00 |
| 13 | 24/01/99 | -52 | 7 | 23:00 | 44 | 23/09/01 | -55 | 5 | 16:00 |
| 14 | 18/02/99 | -123 | 6 | 10:00 | 45 | 19/10/01 | -57 | 10 | 22:00 |
| 15 | 01/03/99 | -94 | 6 | 01:30 | 46 | 24/11/01 | -221 | 10 | 17:00 |
| 16 | 29/03/99 | -56 | 3 | 15:00 | 47 | 21/12/01 | -67 | 8 | 23:00 |
| 17 | 17/04/99 | -90 | 5 | 04:00 | 48 | 30/12/01 | -58 | 3 | 06:00 |
| 18 | 31/07/99 | -53 | 4 | 02:00 | 49 | 10/01/02 | -51 | 10 | 22:00 |
| 19 | 22/09/99 | -173 | 3 | 24:00 | 50 | 02/02/02 | -86 | 7 | 10:00 |
| 20 | 22/10/99 | -237 | 7 | 07:00 | 51 | 01/03/02 | -71 | 5 | 02:00 |
| 21 | 11/01/00 | -81 | 6 | 22:30 | 52 | 17/04/02 | -98 | 6 | 18:00 |
| 22 | 23/01/00 | -97 | 7 | 01:00 | 53 | 11/05/02 | -110 | 6 | 20:00 |
| 23 | 06/04/00 | -287 | 6 | 23:00 | 54 | 19/05/02 | -58 | 4 | 07:00 |
| 24 | 16/04/00 | -60 | 9 | 06/00 | 55 | 01/08/02 | -51 | 3 | 14:00 |
| 25 | 24/04/00 | -61 | 4 | 15:00 | 56 | 04/09/02 | -109 | 3 | 06:00 |
| 26 | 17/05/00 | -92 | 5 | 06:00 | 57 | 24/10/02 | -69 | 5 | 06:00 |
| 27 | 24/05/00 | -147 | 8 | 09:00 | 58 | 20/11/02 | -87 | 4 | 21:00 |
| 28 | 08/06/00 | -90 | 4 | 20:00 | 59 | 27/11/02 | -64 | 8 | 07:00 |
| 29 | 26/06/00 | -53 | 8 | 12:00 | 60 | 19/12/02 | -71 | 7 | 19:00 |
| 30 | 15/07/00 | -289 | 6 | 22:00 | Mean transient | | ~ 6 hours | | |
| 31 | 12/09/00 | -73 | 10 | 20:00 | Number of storms | | 60 | | |